\documentclass[epj]{svjour}
\usepackage{graphicx}

\begin{document}

\title{Infrared spectroscopy of diatomic molecules - a fractional calculus approach}

\author{Richard Herrmann\inst{} 
\thanks{\emph{email address:} herrmann@gigahedron.com}%
}                     
\institute{GigaHedron, Berliner Ring 80, D-63303 Dreieich, Germany}
\date{Received: {\today} / Accepted: {\today}}
\abstract{
The eigenvalue spectrum of the fractional quantum harmonic oscillator is calculated
numerically  solving the fractional Schr\"odinger equation based on the Riemann and Caputo definition of a fractional derivative. The fractional approach allows a smooth transition between vibrational and rotational type spectra, which is shown to be an appropriate tool to analyze IR spectra of diatomic molecules. 
} 
\maketitle
\section{Introduction}
Besides Newton's {\it Principia} Huygens' treatise on the pendulum clock\cite{huy73}   may be considered as one of the most influential works in physics. Since then, the  description of periodical motion  in terms of the classical harmonic oscillator is a standard example for an exactly solvable problem in classical mechanics.  

In non-relativistic quantum mechanics an arbitrary potential may be approximated using a harmonic expansion around the equilibrium position.  Therefore it is the ideal tool to model vibrational degrees of freedom in many different quantum systems. 

Consequently one of the first applications of the quantum harmonic oscillator was the analytic description of the vibrational energy contribution in quantum mechanical models of diatomic molecules \cite{her51}, \cite{whi09}.

To establish the standard model for a successful description of excitation spectra of diatomic molecules in addition to vibrations, rotational degrees of freedom  have to be considered,  so that the total energy of such a system is  given in the simplest approximation mainly as a sum of both contributions (including e.g. anharmonic corrections and a rot-vib interaction term):
\begin{equation}
E_{\rm tot} = E_{\rm vib} + E_{\rm rot}   
\end{equation}
It is the basic intention of this article, to propose an alternative approach based on fractional calculus,  which  overcomes the traditional distinction between rotational and vibrational degrees of freedom by introducing a more universal description, which treats rotations and vibrations simultaneously within the  generalized framework of  fractional oscillations. 

The fractional calculus \cite{old76}-\cite{her11} provides a set of axioms and methods to extend the standard
derivative definition in a reasonable way from integer order n to arbitrary order $\alpha$:
\begin{equation}
 {\partial^n \over \partial x^n}  
\rightarrow
 {\partial^\alpha \over \partial x^\alpha} 
\end{equation}
and therefore allows for an extended,  smooth derivative definition, which may be used to determine generalized symmetries, which go beyond the $U(1)$-symmetry of the standard harmonic oscillator.

In the following, we will first present the main results of a numerical solution of the fractional Schr\"o\-di\-nger equation with the fractional harmonic oscillator potential. 

We will then propose a fractional analogue to the standard rot-vib model used for a description of spectra of diatomic molecules. As a first application, we will use this model to  describe the IR-spectrum of  hydrogen chloride H$^{35}\!$Cl.

\section{The fractional derivative}

We will investigate the spectrum of the fractional quantum harmonic oscillator  for two different definitions
of the fractional derivative, namely the Riemann \cite{rie47} and Caputo \cite{cap67} fractional derivative. Both types are strongly
related.

Starting with the definition of the fractional Riemann integral 
\begin{eqnarray}
{_\textrm{\tiny{R}}}I^\alpha & &  f(x) = \\
 & & 
\cases{
({_\textrm{\tiny{R}}}I_{+}^\alpha f)(x) =  
\frac{1}{\Gamma(\alpha)}   
     \int_{0}^x  d h \, (x-h)^{\alpha-1} f(h)& $x \geq 0$\cr
   ({_\textrm{\tiny{R}}}I_{-}^\alpha f)(x) =  
\frac{1}{\Gamma(\alpha)}   
     \int_x^0  d h \, (h-x)^{\alpha-1} f(h)&   $x<0$
}\nonumber
\end{eqnarray} 
the fractional Riemann derivative is defined as the product of a fractional integration followed by an
ordinary differentiation:
\begin{equation}
\label{dr}
{_\textrm{\tiny{R}}}\partial_x^\alpha =  \frac{\partial}{\partial x}  {_\textrm{\tiny{R}}}I^{1-\alpha}
\end{equation} 
The Caputo definition of a fractional derivative follows an inverted sequence of operations (\ref{dr}).
An ordinary differentiation is followed by a fractional integration
\begin{equation}
{_\textrm{\tiny{C}}}\partial_x^\alpha =   {_\textrm{\tiny{R}}}I^{1-\alpha} \frac{\partial}{\partial x} 
\end{equation}   
Applied to  a function set $f(x)=x^{n \alpha}$ using the Riemann fractional derivative definition (\ref{dr}) we
obtain:
\begin{equation}
\label{xdr}
{_\textrm{\tiny{R}}}\partial_x^\alpha \, x^{n \alpha}  = \frac{\Gamma(1+n \alpha)}{\Gamma(1+(n-1)\alpha)} \, x^{(n-1)\alpha}\quad x \geq 0
\end{equation} 
while for the Caputo definition of the fractional derivative it follows for the same function set:
\begin{eqnarray}
{_\textrm{\tiny{C}}}\partial_x^{\alpha} \, x^{n \alpha} = 
\cases{
\frac{\Gamma(1+n \alpha)}{\Gamma(1+(n-1)\alpha)} \, x^{(n-1)\alpha}&$n > 0$ \cr
0&$n=0  \quad x \geq 0$ 
}
\end{eqnarray} 
Thus for e.g. polynomials with $x^{n \alpha}$ both derivative definitions only differ in the case $n=0$.

\section{Fractional quantum mechanics}
The transition from classical mechanics to quantum mechanics  may be interpreted as a transition from independent coordinate space and momentum space to a Hilbert space, in which space and momentum operators are treated similarly. 

Consequently one postulate of quantum mechanics sta\-tes, that derived results must be independent of the specific choice of e.g. a space or momentum representation.  This is the mathematical manifestation of wave-particle-duality: A description in terms of  either  position $\vec{x}$ or wave vector $\vec{k}$  covers the properties of the same quantum object equivalently.

This implies, that a successful fractional extension of quantum mechanics has to treat coordinates and conjugated momenta equivalently:
\begin{equation}
\label{c8q}
\{x^n, {d^n \over dx^n }\} \rightarrow \{x^\alpha, {d^\alpha \over dx^\alpha }\} \quad \, x>0
\end{equation}
This simultaneous treatment is the major difference between a classical and a quantum mechanical treatment. Furthermore this is the key information needed to quantize any classical quantity like a Hamilton function.

There are several different quantization approaches, like canonical quantization \cite{dir30}-\cite{her05}, path-integrals \cite{fey49}, \cite{las02},  Weyl phase-space quantization \cite{wey27}, \cite{tar08} or stochastic quantization \cite{par81}-\cite{lim08}.

To emphasize the equivalence between a space and a  momentum representation, we will use the classical procedure, proposed by Dirac: 
The classical canonically conjugated 
observables $x$ and $p$ are replaced by a pair of quantum mechanical observables
$\{\hat{x},\hat{p}\}$, which are introduced  as derivative operators on a Hilbert
space of square integrable wave functions $f$. The space  representations 
of these operators are:
\begin{eqnarray}
\label{classic1}
\hat{x} f(x) &= x f(x) \\
\label{classic2}
\hat{p} f(x) &= -i \hbar \partial_x f(x)
\end{eqnarray}
In order to extend this approach to the fractional case, we first make the following statement on parity conservation:
 
The parity properties of the standard coordinate $ x$ and derivative $\partial_x$  should be conserved in the fractional case too. Therefore we postulate the following extensions valid on $R$:
\begin{eqnarray}
\hat{x}^\alpha & = & {\textrm{sign}}(x) |x|^\alpha \\
\hat{D}_x^\alpha & = & {\textrm{sign}}(x)\, _{\textrm{\tiny{R,C}}}\partial_{|x|}^\alpha
\end{eqnarray} 
and introduce the fractional pair  $\{\hat{X},\hat{P}\}$  of  conjugated fractional derivative operators, which are given in space representation:
\begin{eqnarray}
\label{f1}
  \hat{X} \,f(\hat{x}^\alpha)&=&  \left( \frac{\hbar}{m c} \right)^{(1-\alpha)} \hat{x}^\alpha  \,f(\hat{x}^\alpha)\\
\label{f2}
 \hat{P} \,f(\hat{x}^\alpha)&=& -i \left( \frac{\hbar}{m c} \right)^{\alpha} m c \, \hat{D}_x^\alpha  \,f(\hat{x}^\alpha)
\end{eqnarray}
The attached factors $(\hbar/m c)^{(1-\alpha)}$ and $(\hbar/m c)^\alpha m c$ 
ensure correct length and momentum units.
For the special case  $\alpha=1$ 
these definitions correspond to the classical limits (\ref{classic1}) and (\ref{classic2}).

With these operators, we may quantize the classical Hamilton function of the harmonic oscillator:

\section{The fractional quantum harmonic oscillator}
The classical Hamilton function of the harmonic oscillator is given by
\begin{equation}
H_{\rm{class}} = {p^2 \over 2 m} + \frac{1}{2} m \omega^2 x^2
\end{equation}
Following the canonical quantization procedure we replace the
classical observables
$\{x,p \}$
by the fractional derivative operators
 $\{\hat{X},\hat{P}  \}$ according to (\ref{f1}) and (\ref{f2}).
The quantized Hamilton operator $H^\alpha$ results:
\begin{equation}
H^\alpha= {\hat{P}^2 \over 2 m} + \frac{1}{2} m \omega^2 \hat{X}^2
\end{equation}
The stationary Schr\"odinger equation is given by
\begin{eqnarray}
&& H^\alpha \Psi = \nonumber \\
&&  \bigl( - {1 \over 2 m} 
\left( \frac{\hbar}{m c} \right)^{2 \alpha}  \!\! \!\!\!\!\! m^2 c^2  \hat{D}^\alpha_{x}\hat{D}^\alpha_{x}
+ \frac{1}{2}m \omega^2 \left( \frac{\hbar}{m c} \right)^{2(1-\alpha)} \!\!\!\!\!  \!\!\!\!\! |x|^{2 \alpha}
\bigr) \Psi\nonumber \\
 &&=  E'  \Psi
\end{eqnarray}
Introducing the variable $\xi$ and the scaled energy $E$:
\begin{eqnarray}
\xi^\alpha  &=& \sqrt{\frac{m \omega}{\hbar}} 
\left( \frac{\hbar}{m c} \right)^{1- \alpha} x^\alpha \\
E' & = & \hbar \omega E
\end{eqnarray}
we obtain the stationary Schr\"odinger equation for the fractional harmonic oscillator in the  canonical form \cite{her11}:
\begin{equation}
\label{fho}
H^\alpha \Psi_n(\xi) = {1 \over 2 }\bigl( - \hat{D}^{2 \alpha}_{\xi}+ |\xi|^{2 \alpha}
\bigr)\Psi_n(\xi)
 =  E(n,\alpha)  \Psi_n(\xi)
\end{equation}
In contrast to the standard quantum harmonic oscillator \cite{gre09}, the quantum harmonic oscillator  in hyperspherical coordinates for arbitrary integer dimension  \cite{erd53}, \cite{her89} and the radial part of the Schr\"odinger equation for the quantum harmonic oscillator in fractional space dimension respectively \cite{eid11},  the Schr\"odinger equation (\ref{fho})  containing fractional derivatives  has not been solved analytically until now.
\begin{table}[t]
\caption{Parameters in the $_{\textrm{\tiny{R,C}}}\Psi_n^\pm(\xi,\alpha)$ series expansion (\ref{hos1})  and validity ranges of the fractional parameter $\alpha$, which fulfills the requirement of normalizability of the wave function. 
 }
{\begin{tabular}{ll|llll}
\hline\noalign{\smallskip}
type& parity & $\tau$ & $\pi$ & normalizable& $\Psi(\xi\approx 0)$ \\
\noalign{\smallskip}\hline\noalign{\smallskip}
Riemann & even & $\alpha-1$& 0 &  $0.5 \le \alpha \leq 2$ & $ o(\xi^{\alpha-1})$\\
Riemann & odd & $\alpha-1$& $\alpha$ & $0.25 \le \alpha \leq 2$ & $ o(\xi^{2 \alpha-1})$\\
Caputo & even & $0$& 0 &  $0 \le \alpha \leq 2$ & $o(\xi^0)$\\
Caputo & odd & 0& $\alpha$ &  $0 \le \alpha \leq 2$ &  $o(\xi^{\alpha})$\\
\end{tabular}}
\label{tabparms}
\end{table}

An approximate solution for the energy levels has been derived by Laskin  \cite{las02} within the framework of WKB-approxi\-ma\-tion, which is independent from a specific choice of a fractional derivative type:
\begin{equation}
\label{ehowkb}
E_{\textrm{\tiny{WKB}}}(n,\alpha)  =  
\bigl(
n + {1 \over 2}
\bigr)^\alpha
\pi^{\alpha/2}\left({\alpha \Gamma(\frac{1+\alpha}{2 \alpha}) \over \Gamma(\frac{1}{2 \alpha})}\right)^\alpha  
\, n=0,1,2,...
\end{equation}
In a previous work  \cite{he07} we have already emphasized, that these levels allow for a smooth transition from vibrational to rotational types of spectra, depending on the value of the fractional derivative coefficient $\alpha$. 
\begin{eqnarray}
\label{ewkbapprox}
E_{\textrm{\tiny{WKB}}}(n,\alpha \approx 1) & \sim&  
n + {1 \over 2}  \quad   \quad    \quad  n=0,1,2,...\\
E_{\textrm{\tiny{WKB}}}(n,\alpha \approx 2) & \sim&  
 (n+{1 \over 2})^2 =  n (n+1) + 1/4
\end{eqnarray}
The multiplicity for both vibrational and rotational energy levels is 1. 
This implies that an additional symmetry restriction applies  for rotational type degrees of freedom as a consequence of the fractional approach. 
\begin{figure}[t]
\begin{center}
\includegraphics[width=88mm]{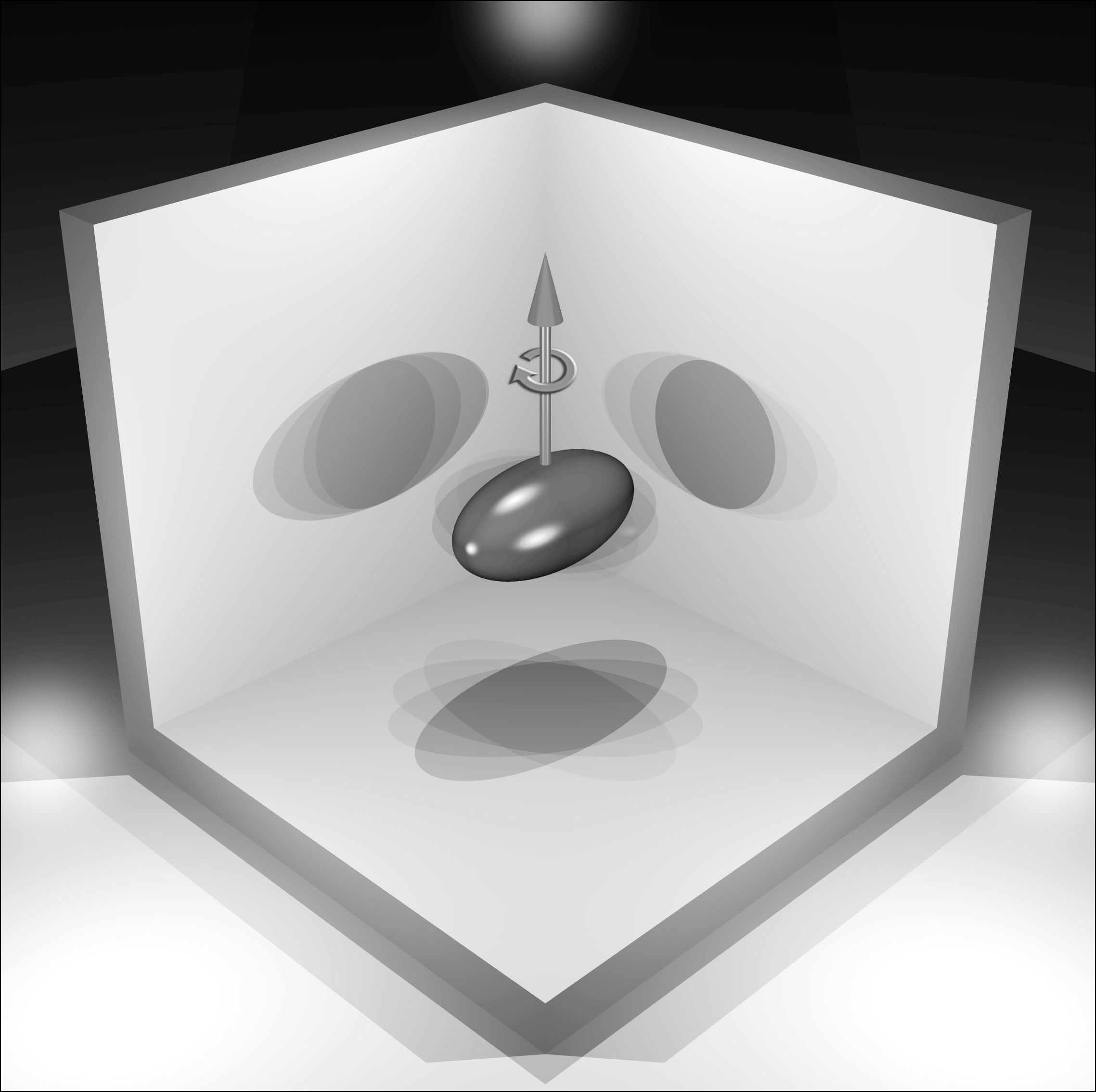}\\
\caption{\label{figrotvibHO}
Geometric interpretation of the group transmutation  $  SO(3)\rightarrow_{\textrm{\tiny{sym}}}\!\!\!SO(3) \rightarrow U(1)$ in terms of a dimensional reduction: 
\newline
The rotational state  $ |L M \rangle$, characterized by two quantum numbers $L,M$ may be  projected from 3-dimensional space onto the 2-dimensional  x-y plane by fixing the rotation axis as a consequence of application of  the additional symmetry requirement $M=+L$. \newline
This may then be interpreted as a vibrational degree of freedom with a fixed  phase relation \cite{lis57}  in both x and y direction respectively, where the state $|n  \rangle \equiv |n n  \rangle$ is  determined by a single quantum number  $n=L$. 
\newline
The same argument holds for the  inverse transition $  SO(3)\leftarrow_{\textrm{\tiny{sym}}}\!\!\!SO(3) \leftarrow U(1)$, see discussion  of (\ref{go})ff.  The fractional derivative coefficient  $\alpha$ serves as an order parameter for a smooth transition between these cases.} 
\end{center}
\end{figure}

In order to give a geometric interpretation,
let us recall, that for $\alpha=2$ in the classical approach rotational states $ |L M \rangle$ are classified  according to the group chain $SO(3) \supset SO(2)$ by the two quantum numbers $L,M$. Thus the additional symmetry  may  be imposed as a geometric constraint  by fixing the rotation axis in space setting  e.g. $M=+L$, see also \cite{he08}, \cite{he09}. 

This additional symmetry requirement  reduces the multiplicity of a $SO(3)$ multiplet for given $L=n$ from $2 n+1$ to 1 and the corresponding rotational states $|n n \rangle$ are characterized by a single  quantum number $n$, see figure \ref{figrotvibHO}. 

\begin{figure}[t]
\begin{center}
\includegraphics[width=88mm]{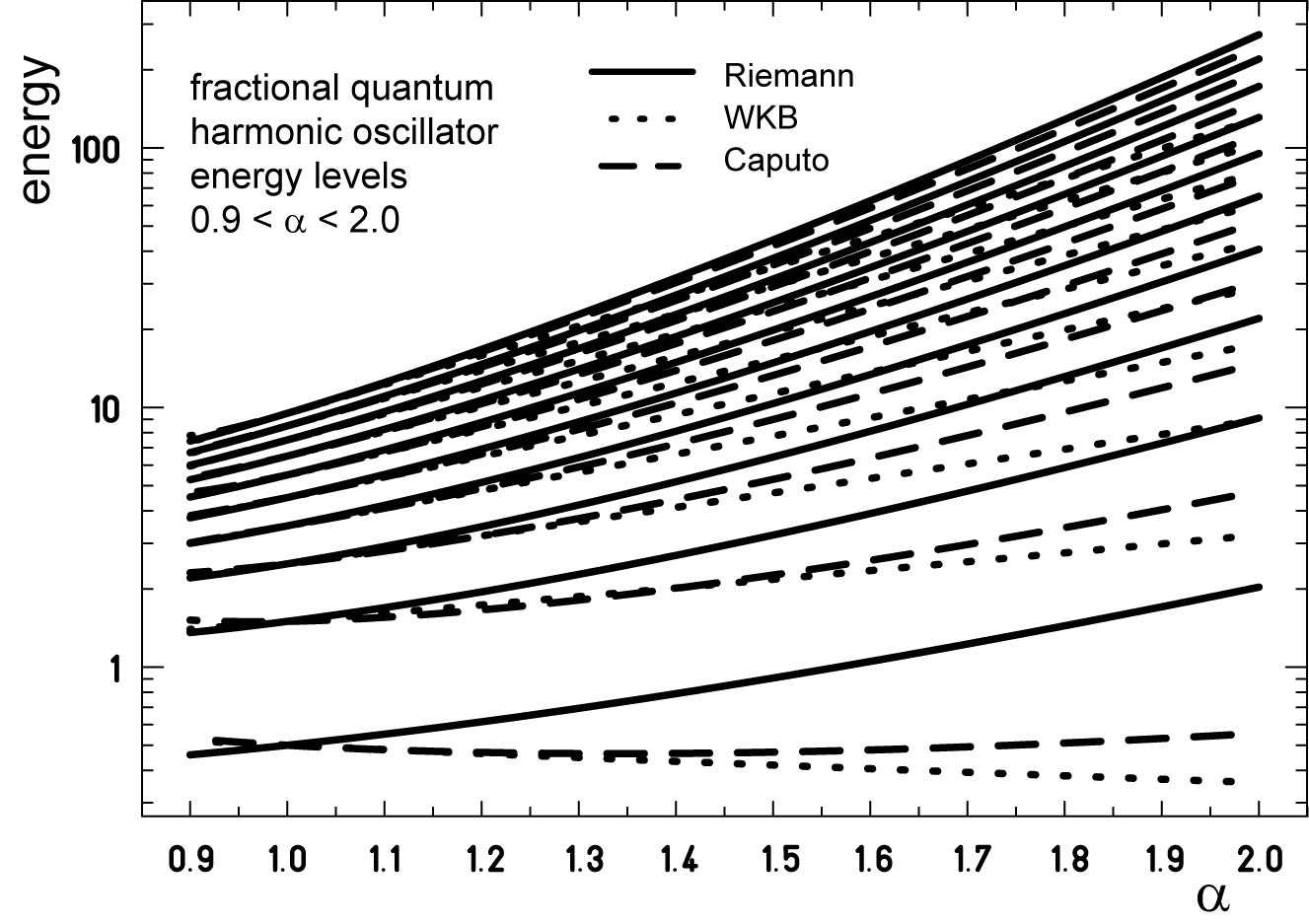}\\
\caption{\label{eho1to2}
Logarithmic plot of the first 10 energy levels of the fractional quantum harmonic oscillator in the range $0.90 \leq \alpha \leq 2$ based on the Riemann (thick lines), Caputo (dashed lines) and WKB-approximation (dotted lines) of the fractional derivative definition.
} 
\end{center}
\end{figure}

\begin{figure}[t]
\begin{center}
\includegraphics[width=88mm]{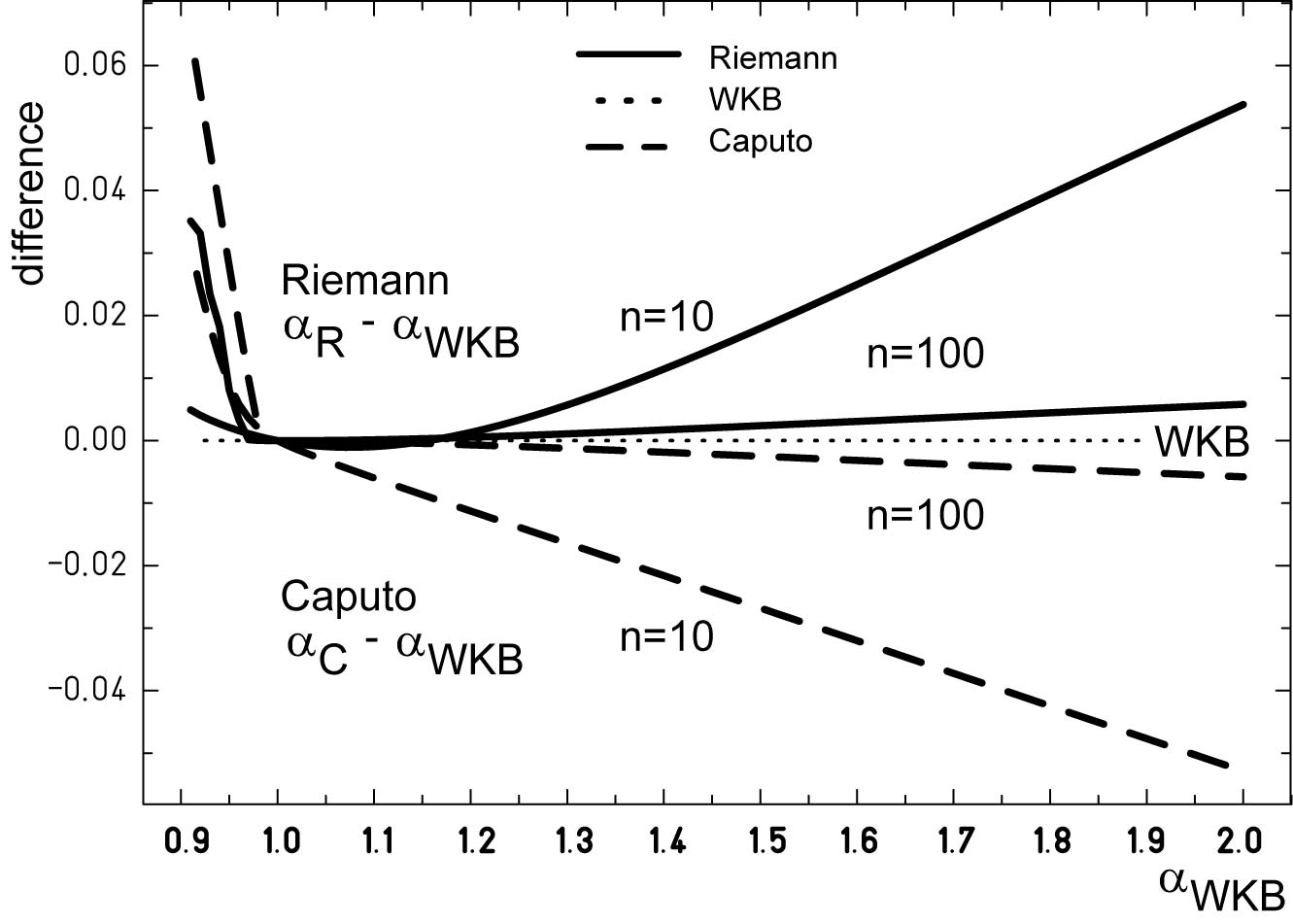}\\
\caption{\label{compare}
Result of a least square fit of energy level spectra based on the Riemann (thick lines)  and Caputo (dashed lines)  derivative definition with the WKB-approximation  (dotted line)  (\ref{ehowkb}). Plotted is the difference of $\alpha$ values for a given number of energy levels ($n=10,100$) with fitted $\alpha_{\textrm{\tiny{WKB}}}$.  
} 
\end{center}
\end{figure}

Hence  the symmetry of the fractional quantum harmonic oscillator bridges two different Lie-algebras, 
it is the transmutation  of one group  into another, an idea, which also motivated the development of q-deformed Lie-algebras \cite{he07}-\cite{har06}. 

Therefore the fractional harmonic oscillator may be a useful tool to describe rot-vib spectra, with $\alpha$ in the range of $1 \leq \alpha \leq 2$. The fractional harmonic oscillator could be of similar importance as its classical counterpart, an exact solution is of fundamental interest.

The main results from a numerical solution of the cano\-nical Schr\"odinger equation of the fractional quantum harmonic oscillator (\ref{fho}) using the Riemann and Caputo fractional derivative definition are presented in the next section. 
\section{Numerical solution}

For a numerical solution of (\ref{fho}), we expand the wave function in a fractional Taylor-series on the positive semi-axis $\xi \geq 0$. Furthermore the wave function should be an eigenfunction of the parity operator $\Pi$:
\begin{eqnarray}
\label{hos1}
_{\textrm{\tiny{R,C}}}\Psi_n^\pm(\xi,\alpha)  &=&  
\xi^{\tau} \xi^\pi \sum_{i=0}^{N}a_i(E) \xi^{2 \alpha i}  \quad \xi> 0 \\
\Pi_\textrm{{\tiny{R,C}}}\Psi_n^\pm(\xi,\alpha)  &=&  
\pm_\textrm{{\tiny{R,C}}}\Psi_n^\pm(\xi,\alpha) 
\end{eqnarray}
where the parameters $\tau$ and $\pi$ determining the type (Riemann or Caputo) of the fractional derivative definition used  and parity of the wave function are listed in table \ref{tabparms}. The $\pm$ index indicates the parity for $n$ even and odd respectively.

The coefficients $a_i(E)$ are determined using a standard shooting method: 

From the requirement, that the wave function should vanish at infinity for a given eigenvalue $E_n$ follows a determining condition for sufficiently large distance $\xi_{\textrm{\tiny{det}}}$:  
\begin{equation}
\label{cond1}
{\lim_{E \rightarrow E_n}} \,_{\textrm{\tiny{R,C}}}\Psi_n^\pm(\xi_{\textrm{\tiny{det}}},\alpha, E)  =  0  
\end{equation}
 which allows for an iterative procedure to calculate $E_n$ with arbitrary precision.

For practical calculations, we determined the eigenvalues with a precision of 32 significant digits with the settings $20 \leq \xi_{det} \leq 40$, $1000 \leq N \leq 10000$, depending on $\alpha$ and $\Delta \alpha=0.01$.

\begin{figure*}
\begin{center}
\includegraphics[width=180mm]{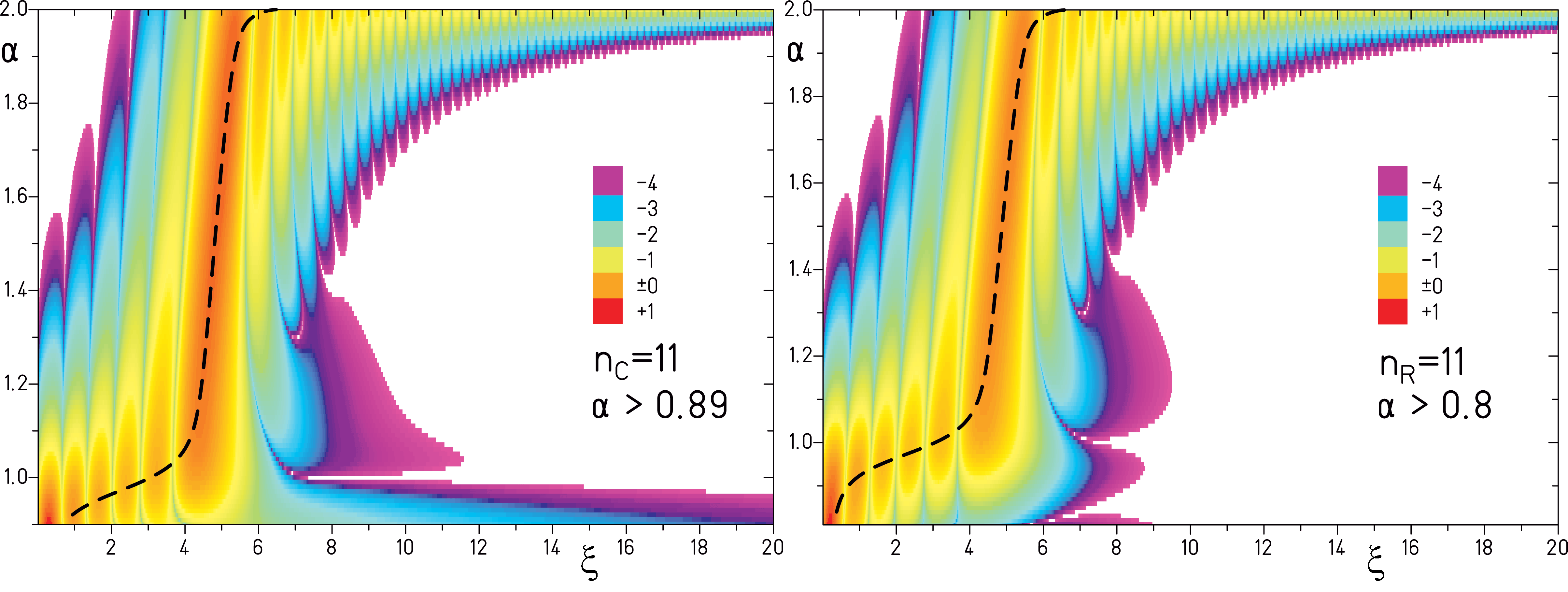}\\
\caption{\label{both}
Logarithmic probability density  $|\Psi_{n=11}^-(\xi,\alpha)|^2$ for the solution of the fractional quantum harmonic oscillator based on the Caputo (left) and Riemann (right)  definition of the fractional derivative in the full range of allowed $\alpha$ values for real energy eigenvalues. The dashed line indicates the expectation value for the modified position operator $\langle \hat{\xi} \rangle$. 
} 
\end{center}
\end{figure*}

The major restriction for the range of allowed $\alpha$ values is the requirement of normalizability of the wave function obtained. 
From 
\begin{equation}
\label{cond2}
\int_{\infty}^\infty d\xi _\textrm{{\tiny{R,C}}}\Psi_n^\pm(\xi,\alpha) _\textrm{{\tiny{R,C}}}\Psi_n^\pm(\xi,\alpha)^* < \infty
\end{equation}
follows an upper bound $\alpha \leq 2$. Of course, for $\alpha > 2$ condition (\ref{cond1}) may still be used as an equivalent of a box-normalization condition, but the results are no solutions of the harmonic oscillator potential any more.

In figure \ref{eho1to2} we compare the resulting energy level spectra to the WKB-approximation. While in the vicinity of $\alpha=1$ differences are negligible, for larger $\alpha$ values especially for low lying energies the calculated spectra differ  significantly, but as a first remarkable result we obtain similar to the case of WKB-approximation the expected  smooth transition from vibrational to rotational type of spectrum using  the Riemann and Caputo fractional derivative definition respectively.

The results of a fit of the energy levels with the WKB-approximation are plotted in figure \ref{compare}. We may deduce, that for increasing number of energy levels  $n$ eq. (\ref{ehowkb}) approximates the energy values better. From this figure we may also deduce, that rotational spectra, which are characterized by $\alpha_{\textrm{\tiny{WKB}}}=2$ are reproduced using the Caputo derivative for a slightly smaller $\alpha$ value,  $\alpha_{\textrm{\tiny{C}}}<2$, while for the Riemann derivative we have to use a slightly larger $\alpha$ value, $\alpha_{\textrm{\tiny{R}}}>2$ and therefore using the Riemann fractional derivative definition implies the need for box normalization, if we want to generate the exact level spacing of rotational type.

The behavior of the eigenfunctions $\Psi_n^\pm(\xi,\alpha)$ differs significantly in the region $1 \leq \alpha \leq 2$ and $\alpha < 1$. As an example, in figure \ref{both} we have plotted the occupation probability for the eigenfunction $\Psi_{n=11}^-(\xi,\alpha)$ for the Caputo  and Riemann solution. 

Introducing the modified position operator $\bar{\xi}=\theta(\xi) \xi$, its expectation value
\begin{equation} 
\label{pos}
\langle \bar{\xi} \rangle = 
\int_{0}^\infty d\xi _\textrm{{\tiny{R,C}}}\Psi_n^\pm(\xi,\alpha) \, \xi \, _\textrm{{\tiny{R,C}}}\Psi_n^\pm(\xi,\alpha)^* 
\end{equation} 
yields the position information on the positive semi-axis.   This value is almost constant for $\alpha>1$. For increasing $\alpha$ only the side maxima are changing from inside to outside and therefore mark a smooth transition from Hermite-type  to Bessel-type polynomials.

In contrast for $\alpha<1$, the position value  tends very fast  to zero which means, that the wave function becomes strongly located at the origin in this case.

As a consequence, the energy diagram for $\alpha<1$ is dominated by the kinetic term, since for the potential near the origin
\begin{equation} 
\label{pot}
\lim_{\xi \rightarrow 0}|\xi|^{2 \alpha} = 0 
\end{equation} 
holds.

This means, that the solutions of the fractional Schr\"odi\-nger equation for the harmonic oscillator and the solutions for the free  fractional Schr\"odinger equation more and more coincide for $\alpha<1$, a behavior, which we expect for any potential, which vanishes at the origin.

The eigenfunctions of the free fractional Schr\"odinger equation are known analytically and are given in terms of the Mittag-Leffler functions $E_\alpha(z)$ and $E_{\alpha, \beta} (z)$ as \cite{her11}, \cite{her05}:
\begin{eqnarray} 
\label{pot0}
\textrm{{\tiny{C}}}\Psi_n^{+\textrm{\tiny{free}}}(\xi,\alpha) &=&  E_{2 \alpha}( -\xi^{2 \alpha})\\
\textrm{{\tiny{C}}}\Psi_n^{-\textrm{\tiny{free}}}(\xi,\alpha) &=&  \xi^\alpha E_{2\alpha,1+\alpha}( -\xi^{2 \alpha})\\
\textrm{{\tiny{R}}}\Psi_n^{+\textrm{\tiny{free}}}(\xi,\alpha) &=&  \xi^{\alpha-1}E_{2 \alpha,\alpha}( -\xi^{2 \alpha})\\
\textrm{{\tiny{R}}}\Psi_n^{-\textrm{\tiny{free}}}(\xi,\alpha) &=&  \xi^{2 \alpha-1}E_{2\alpha,2 \alpha}( -\xi^{2 \alpha})
\end{eqnarray} 
and the corresponding eigenvalues are determined  from the zeros of these functions. For $\alpha<1$ there exists only a finite number of zeros and therefore the energy spectrum is limited \cite{her05}, \cite{hil06}-\cite{sey08}.

\begin{figure}
\begin{center}
\includegraphics[width=88mm]{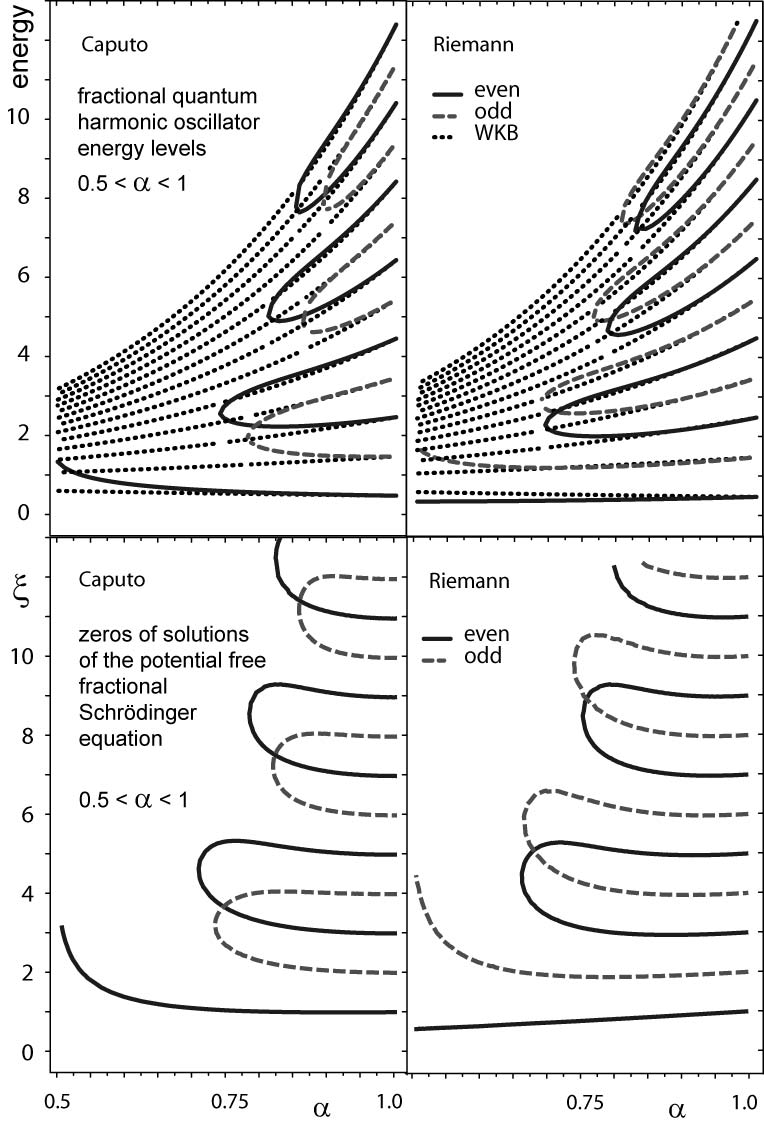}\\
\caption{\label{lev2}
In the upper row energy levels in the range $0.5< \alpha<1$ for the Caputo (left column) and Riemann (right column) fractional quantum harmonic oscillator are plotted.  Thick lines indicate  positive parity, dashed lines indicate odd parity of the corresponding solution. For decreasing $\alpha$, the number of real eigenvalues is reduced.  Dotted lines show the WKB-approximation $E_{\textrm{\tiny{WKB}}}$. 
\newline
The lower row shows the zeros of eigenfunctions for the potential free fractional Schr\"odinger equation \cite{her11}, \cite{her05}. 
} 
\end{center}
\end{figure}

In the upper row of figure \ref{lev2} we show the lowest energy levels for the fractional harmonic oscillator for $\alpha<1$. This spectrum is limited to a finite number of levels, which may now be understood  in comparison to the plot of zeros of the solutions of the free fractional Schr\"odinger equation presented in the lower row of figure   \ref{lev2}.  

It can be deduced, that the onset of real eigenvalues for a given $\alpha$ of the fractional  harmonic oscillator and the occurrence of a zero in the potential free solution agree qualitatively.

Since the free fractional Schr\"odinger equation is formally equivalent to the classical fractional harmonic oscillator differential equation we may conclude, that in the special case of the harmonic oscillator the transition from fractional classical mechanics to fractional quantum mechanics is smooth.

We have thus demonstrated, that the fractional quantum harmonic oscillator extends the symmetry of the standard harmonic oscillator and allows for a generalized description of rotational and vibrational degrees of freedom.

Since we want to emphasize this unifying point of view, an appropriate area of application is molecular spectros\-copy, where an interplay between vibrations and rotations is observed. 

Therefore we will investigate the properties of infrared spectra of diatomic molecules in the next section using the fractional quantum harmonic oscillator. 

\section{Application: The infrared-spectrum of  H$^{35}\!$Cl}

Traditionally diatomic molecules are treated in lowest order as rigid rotors with a fixed bond length, which in addition are able to perform vibrations around this equilibrium position.  

For higher excitations, anharmonic contributions  and the influence of centrifugal stretching are taken into account via a series expansion, which we present here up to third order:
\begin{eqnarray}
\label{Erv}
E_{\textrm{\tiny{rot-vib}}}&&(\nu, J)  =
E_{\textrm{\tiny{rot}}} +E_{\textrm{\tiny{vib}}} =  \\
&&B_\nu J(J+1) + D_\nu J^2(J+1)^2   + \nonumber\\
&&\hbar \omega_e (\nu+\frac{1}{2}) + \hbar \omega_e x_e (\nu+\frac{1}{2})^2 + \hbar \omega_e y_e (\nu+\frac{1}{2})^3   \nonumber  \end{eqnarray}
where the constants $B_\nu$ and $D_\nu$ add a rot-vib coupling via:
\begin{eqnarray}
B_\nu  = B_e - \alpha_e (\nu + \frac{1}{2}) \\     
D_\nu  = D_e - \beta_e (\nu + \frac{1}{2})   
\end{eqnarray}
which results in a seven parameter energy formula to determine the energy levels in the rot-vib model.

Now we introduce the  fractional quantum harmonic oscillator model, which describes rotational and vibrational spectra from a unifying point of view.
As we pointed out in the previous section, depending on the value of the fractional derivative parameter $\alpha$  we may generate both types of spectra:
\begin{eqnarray}
E_{\textrm{\tiny{rot}}}(J)  &=& E(\alpha \approx 2,J) \\
E_{\textrm{\tiny{vib}}}(\nu)  &=& E(\alpha \approx 1,\nu)
\end{eqnarray}
Thus we propose the fractional analogue of the standard rot-vib-model:
\begin{eqnarray}
\label{Ervfrac}
E^{\textrm{\tiny{fractional}}}_{\textrm{\tiny{rot-vib}}}&&(\nu, J)  = \nonumber \\
&&E_{\textrm{\tiny{rot}}} +E_{\textrm{\tiny{vib}}} +E_{\textrm{\tiny{coupling}}} = \nonumber \\
&&\tilde{B}_e E(\alpha_J \approx 2, J) + 
\hbar \tilde{\omega}_e E(\alpha_\nu \approx 1, \nu)+\nonumber \\
&&\tilde{\alpha}_e   E(\alpha_J \approx 2, J) E(\alpha_\nu \approx 1, \nu)+g_0
\end{eqnarray}
with six parameters $\alpha_J$, $\tilde{B}_e$,  $\alpha_\nu$,  $\hbar \tilde{\omega}_e$, $\tilde{\alpha}_e$, $g_0$.
A least square fit of the fractional rot-vib model energies with the standard rot-vib model energies  yields the parameter sets listed in table \ref{tab2}. 
\begin{table}[t]
\caption{Parameter sets  to determine the energy levels of H$^{35}\!$Cl for the standard rot-vib-model (\ref{Erv}) in the first row \cite{her51} and for the proposed generalized fractional model (\ref{Ervfrac}) for WKB, Riemann and Caputo definition of the fractional derivative. Finally the rms-error is given.}
{\begin{tabular}{l|llll}
\hline\noalign{\smallskip}
stan- &$B_e$& $\alpha_e$ & $D_e$ &  $\beta_e$ \\
dard&$10.59341$ & $0.30718$ & $5.3194*10^{-4}$&$7.51*10^{-6}$\\
&$\hbar \omega_e$& $\hbar \omega_e x_e$ &$\hbar \omega_e y_e$&rms-error\\
&  $2990.946$ & $ 52.8186$&$0.2243$&$3.8\%$\\
\noalign{\smallskip}\hline\noalign{\smallskip}
\noalign{\smallskip}\hline\noalign{\smallskip}
WKB&$\tilde{B}_e$& $\tilde{\alpha}_e$&$g_0$&  $\hbar \tilde{\omega}_e$\\
& $7.406$&$0.202$ &$152.016$& $2707.77$\\
& $\alpha_J$&  $\alpha_\nu$& &rms-error\\
& $1.997$ & $1.051$& &$3.3\%$ \\
\noalign{\smallskip}\hline\noalign{\smallskip}
Rie&$\tilde{B}_e$& $\tilde{\alpha}_e$&$g_0$&  $\hbar \tilde{\omega}_e$\\
& $2.588$&$0.080$ &$10.355$& $3095.36$\\
& $\alpha_J$&  $\alpha_\nu$&& rms-error\\
& $2.081$ & $0.937$& &$3.8\%$\\
\noalign{\smallskip}\hline\noalign{\smallskip}
Cap&$\tilde{B}_e$& $\tilde{\alpha}_e$&$g_0$&  $\hbar \tilde{\omega}_e$\\
& $5.264$&$0.155$ &$-64.973$& $2938.77$\\
& $\alpha_J$&  $\alpha_\nu$& &rms-error\\
& $1.91$ & $0.914$& &$3.8\%$ \\
\noalign{\smallskip}\hline\noalign{\smallskip}
\end{tabular}}
\label{tab2}
\end{table}

\begin{figure}
\begin{center}
\includegraphics[width=88mm]{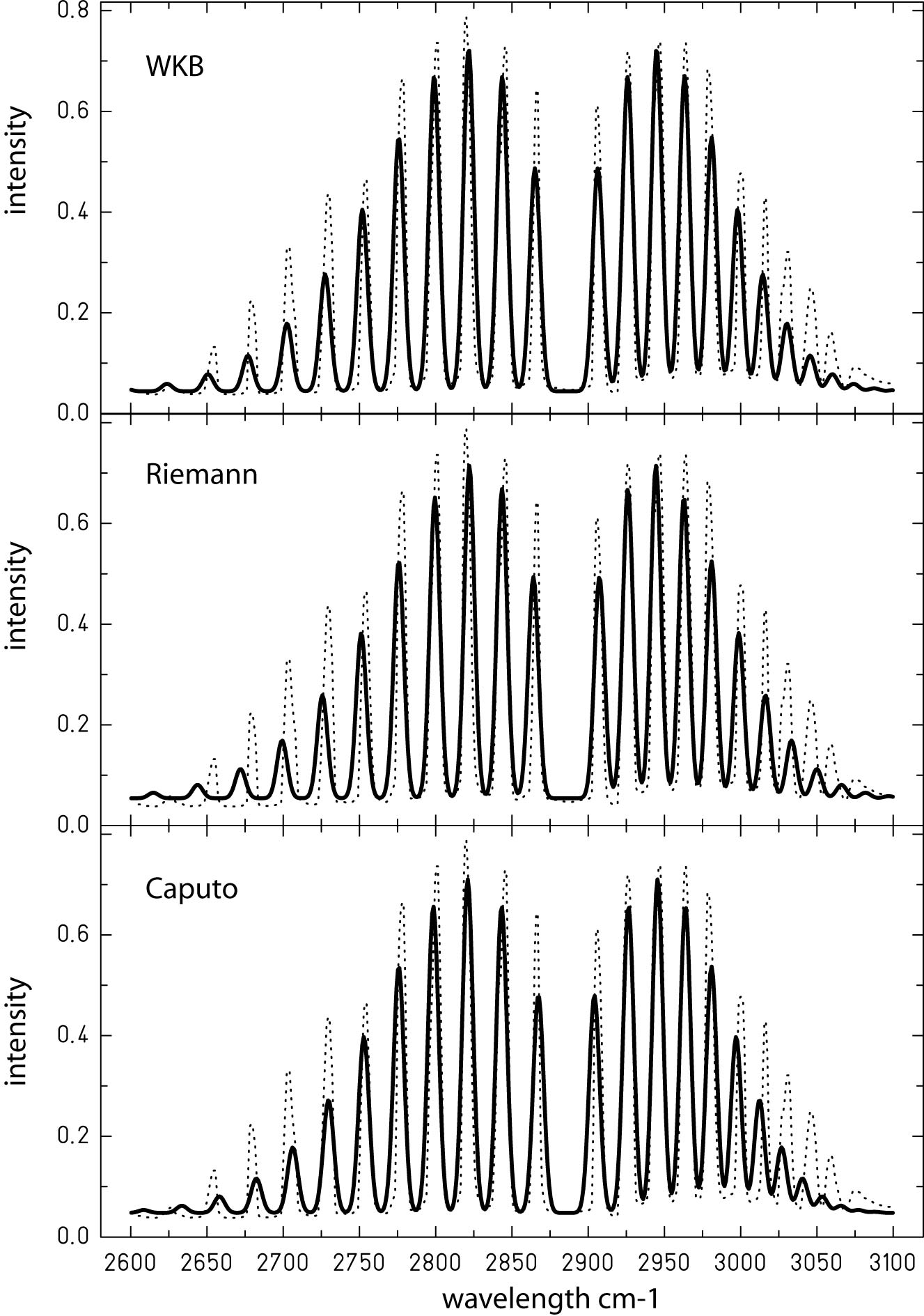}\\
\caption{\label{hclfig}
Fit results for the infrared spectrum of  H$^{35}\!$Cl in the wavelength range from 2600-3100 $[\textrm{cm}^{-1}]$.
Dotted lines are the experimental values from \cite{nis12}. From top to bottom the result for the energy levels of the fractional rot-vib-model (\ref{Ervfrac}) using the WKB-approximation, the Riemann and the Caputo fractional derivative definition.
} 
\end{center}
\end{figure}

In order to emphasize the property of the fractional rot-vib model to realize rotational type spectra, we want to 
explore the lowest energy photon absorption processes in the infrared region. In that case from $\Delta J = \pm1$, we obtain two branches (R and P) with the properties:
\begin{eqnarray}
\label{branch}
\Delta E_{R}(J) = E(\nu=1, J + 1)  - E(\nu=0, J)  \, J=0,...\\
\Delta E_{P}(J) = E(\nu=1, J - 1)  - E(\nu=0, J)  \,  J=1,...
\end{eqnarray}
 These transition energies may be directly compared to the experimental spectrum. The corresponding intensities $I_j$ for a given transition are determined in the standard case by the Boltzmann-distribution
\begin{equation}
I_j = (2 J + 1) e^{-\beta E_{\textrm{\tiny{rot}}}} = (2 J + 1) e^{-B_\nu J (J+1)/kT}
\end{equation}
where $\beta= 1/kT$ and k is the Boltzmann constant. The factor $(2J +1)$ is a result of the multiplicity of a given rotational state. 

In the case of the fractional rot-vib model this implies the following generalization:
\begin{equation}
\label{go}
I_j^{\textrm{\tiny{fractional}}} =m(\alpha_J) e^{-E_{\textrm{\tiny{rot}}}/kT} = (2(\alpha_J-1) J + 1) e^{-E_{\textrm{\tiny{rot}}}/kT}
\end{equation}
with the fractional multiplicity $m(\alpha)$.  For varying $\alpha$ this factor 
\begin{equation}
\label{m}
m(\alpha) = 2(\alpha-1) J + 1 
\end{equation}
may now be interpreted physically as a smooth transition  of multiplicities $m$
from 1 to $2 J + 1$ in the range $1 \leq \alpha \leq 2$; hence  (\ref{m}) indeed  completes the proper transition from $U(1)$ to $SO(3)$ by restoring the isotropy of space.  

Furthermore   we obtain a  physical explanation for the limited number of energy levels in the case $\alpha<1$: 

From the requirement that  the multiplicity  of a given state $J$ should be  a positive real number, $m>0$, which means, within the framework of a possible particle-hole formalism we restrict to a description of particles only,  using (\ref{m}) a condition follows
for the  finite set of allowed J values $J \in \{0,...,J_{\textrm{\tiny{max}}}\}$: 
\begin{equation}
J{\textrm{\tiny{max}}} < {{1} \over {2(1-\alpha)}},  \quad \quad \alpha < 1
\end{equation}
which in the limiting case  $\alpha \rightarrow {{1} \over{2}}$ reduces to only one allowed level with $J=0$. 

This is  in qualitative agreement with  numerical results presented in figure \ref{lev2} and a strong argument for the validity and physical justification of our choice of the Riemann and Caputo definition of the fractional derivative.

These considerations establish a connection between continuous, fractional space dimension D and the fractional quantum harmonic oscillator with order parameter $\alpha$ as
\begin{equation}
D = 2 \alpha-1, \quad \quad {{1} \over{2}}<\alpha \leq 2
\end{equation}
which sets an upper limit to the space dimension $D\le 3$ as a consequence of the requirement of normalizability of the wave function and a lower limit of $D \ge 0$, where we observe a dimensional freezing of vibrational degrees of freedom for a point particle and therefore  gives an alternative answer to Ehrenfest's thought-provoking question  \cite{ehr17} 100 years after; see also \cite{he90}.

With intensities (\ref{go}) and assuming an exponential distribution for every 
transition we are able to fit the experimental spectrum \cite{nis12}. In table \ref{tab2} the optimized parameters are listed. The resulting spectra are presented in figure  \ref{hclfig}. 

The overall error is less than $4 \%$. As a remarkable fact, the experimental spectrum is reproduced in the case of the Riemann fractional harmonic oscillator with $\alpha_J > 2$, which means, the corresponding eigenfunctions are only box-normalizable. In addition, the position of the maxima is slightly deviating from the experiment for large $J$ values. In case of the Caputo derivative definition, these positions are described much better.

Consequently, only the Caputo fractio\-nal rot-vib model describes the experiment within the allowed parameter range.

It is remarkable, that in the case of the fractional rigid rotor \cite{he17}, which is an alternative description of rot-vib systems, just the opposite conclusion was drawn: the Caputo derivative based version was discarded and only the Riemann rotor model was acceptable, since only in this case the zero-point energy contribution of vibrational modes was treated  correctly.

\section{Conclusion and outlook}
It has been demonstrated, that the fractional quantum harmonic oscillator generates an energy spectrum, which, besides vibrational degrees of freedom also shows rotational type spectra. The fractional parameter $\alpha$ allows for a smooth transition between these two extreme cases.

Therefore vibrations and rotations may be treated equi\-va\-lent\-ly in a unified generalized
rot-vib model, which may be successfully used to reproduce the infrared spectra of diatomic molecules, which has been demonstrated for the case of hydrogen chloride.  

It has also been shown, that the WKB-approximation is a useful tool to approximate higher energy levels, but in the low energy region the exact solutions using the Riemann and Caputo derivative definitions differ significantly and cover a much broader area of useful applications.

In this paper we presented first results for the fractional quantum harmonic oscillator. The results encourage further studies in this field, especially the knowledge of eigenvalues and eigenfunctions for higher energy levels  \cite{he13} will be useful to describe highly excited rotational molecular states, an area, which has been made accessible recently, see e.g. \cite{vil00},\cite{yua11}. 

Finally we  collected arguments in support of the idea, that  a promising field of future research is the formulation of a  quantum statistical extension of fractional  thermodynamics \cite{abe97}, \cite {hil00} in terms of the fractional extension of the partition function $Z$ using the fractional generalization of a  Hamiltonian $H^\alpha$ e.g. (\ref{fho}):
\begin{equation}
Z^\alpha = m(\alpha) \textrm{Tr}(e^{-\beta H^\alpha})
\end{equation}
with the additional fractional multiplicity $m(\alpha)$, which emphasizes the connection between fractional calculus and fractional space dimensions. Closer examinations on this subject will be presented in future works.

\section{Acknowledgment}
We thank A. Friedrich, G. Plunien from TU  Dresden, Germany, D. Troltenier, SAP Labs, Exton, PA, USA and M. Ortigueira, FCT, Lisboa, Portugal for useful discussions. Figure \ref{figrotvibHO} was generated using the free 3D-fractal landscape generator Bryce \cite{bry12}.


\end{document}